\documentclass[reprint,superscriptaddress,aps]{revtex4-1}
\usepackage{amsmath}
\usepackage{amssymb}
\usepackage{bm}
\usepackage{braket}
\usepackage{color}
\usepackage[varg]{txfonts}
\usepackage{varwidth}
\usepackage{dcolumn}
\usepackage[breaklinks,colorlinks=true,linkcolor=blue,urlcolor=cyan,citecolor=blue]{hyperref}
\usepackage[dvipdfmx]{graphicx}
\usepackage{lineno}

\begin{document}

\title{Magnetoelastic mapping of the high-field phase diagram in the topological cubic helimagnet SrFeO$_{3}$}

\author{Masaki~Gen}
\email{gen@issp.u-tokyo.ac.jp}
\affiliation{Institute for Solid State Physics, University of Tokyo, Kashiwa 277-8581, Japan}
\affiliation{RIKEN Center for Emergent Matter Science (CEMS), Wako 351-0198, Japan}

\author{Shun~Okumura}
\email{okumura@ap.t.u-tokyo.ac.jp}
\affiliation{RIKEN Center for Emergent Matter Science (CEMS), Wako 351-0198, Japan}
\affiliation{Quantum-Phase Electronics Center, The University of Tokyo, Tokyo 113-8656, Japan}
\affiliation{International Institute for Sustainability with Knotted Chiral Meta Matter (WPI-SKCM$^2$),
Hiroshima University, Hiroshima 739-8531, Japan}

\author{Shusaku~Imajo}
\affiliation{Institute for Solid State Physics, University of Tokyo, Kashiwa 277-8581, Japan}
\affiliation{Department of Advanced Materials Science, The University of Tokyo, Kashiwa 277-8561, Japan}

\author{Taro~Nakajima}
\affiliation{Institute for Solid State Physics, University of Tokyo, Kashiwa 277-8581, Japan}
\affiliation{RIKEN Center for Emergent Matter Science (CEMS), Wako 351-0198, Japan}
\affiliation{Institute of Materials Structure Science, High Energy Accelerator Research Organization, Tsukuba 305-0801, Japan}

\author{Karel~Prokes}
\affiliation{Helmholtz-Zentrum Berlin f\"{u}r Materialien und Energie, Hahn-Meitner Platz 1, 14109 Berlin, Germany}

\author{Shunsuke~Kitou}
\affiliation{Department of Advanced Materials Science, The University of Tokyo, Kashiwa 277-8561, Japan}

\author{Yusuke~Tokunaga}
\affiliation{Department of Advanced Materials Science, The University of Tokyo, Kashiwa 277-8561, Japan}

\author{Yoichi~Nii}
\affiliation{Department of Applied Physics and Physico-Informatics, Faculty of Science and Technology, Keio University, Yokohama 223-8522, Japan}

\author{Koichi~Kindo}
\affiliation{Institute for Solid State Physics, University of Tokyo, Kashiwa 277-8581, Japan}

\author{Yoshimitsu~Kohama}
\affiliation{Institute for Solid State Physics, University of Tokyo, Kashiwa 277-8581, Japan}

\author{Shintaro~Ishiwata}
\affiliation{Division of Materials Physics, Graduate School of Engineering Science, The University of Osaka, Osaka 560-8531, Japan}

\author{Taka-hisa~Arima}
\affiliation{RIKEN Center for Emergent Matter Science (CEMS), Wako 351-0198, Japan}
\affiliation{Department of Advanced Materials Science, The University of Tokyo, Kashiwa 277-8561, Japan}

\begin{abstract}

The cubic perovskite SrFeO$_{3}$ is a prototypical centrosymmetric itinerant magnet that hosts a quadruple-${\mathbf Q}$ hedgehog--antihedgehog lattice and exhibits a complex magnetic-field--temperature phase diagram.
Yet, the microscopic mechanism underlying the emergence of its versatile multiple-${\mathbf Q}$ phases remains unresolved.
Here, we reveal the field-orientation dependence of the magnetic phase diagram and establish an effective spin Hamiltonian for SrFeO$_{3}$ that incorporates a cubic single-ion anisotropy together with bilinear and biquadratic interactions in momentum space, which originate from the spin--charge coupling.
In addition, we observe magnetoelastic signatures of a redistribution of the ligand-hole density upon entering the forced ferromagnetic phase. 
These findings emphasize the pivotal importance of electronic itinerancy arising from the formation of a ligand-hole band in stabilizing multiple-${\mathbf Q}$ phases.

\end{abstract}

\date{\today}
\maketitle

\section{\label{Sec1} Introduction} 
\vspace{-0.2cm}

Noncoplanar magnetic structures with nontrivial topology, exemplified by magnetic skyrmions, have recently attracted tremendous attention owing to their rich emergent electrodynamics and potential applications in spintronic devices \cite{2013_Fer, 2013_Nag, 2021_Tok}.
Magnetic skyrmions are two-dimensional swirling spin textures composed of a superposition of multiple modulation vectors.
They host emergent magnetic fields that give rise to unconventional transport phenomena, such as the topological Hall effect \cite{2009_Neu, 2019_Kur}, and provide a fertile platform for exploring novel collective excitations \cite{2012_Ono, 2020_Sek} and current-driven dynamics \cite{2010_Jon, 2024_Bir}.
Extending such spin textures into three dimensions leads to the so-called hedgehog--antihedgehog lattice (HL), which can be regarded as a periodic array of emergent magnetic monopoles and antimonopoles.
The discovery of triple-${\mathbf Q}$ and quadruple-${\mathbf Q}$ HL phases in chiral magnets MnGe \cite{2012_Kan, 2015_Tan, 2020_Kan} and Mn(Si$_{1-x}$Ge$_{x}$) \cite{2019_Fuj}, respectively, has triggered a surge of theoretical investigations addressing their stabilization mechanisms \cite{2006_Bin_PRL, 2006_Bin_PRB, 2011_Par, 2016_Yan, 2020_Gry, 2020_Oku, 2021_Shi, 2022_Kat, 2022_Oku, 2023_Kat, 2023_Yam, 2026_May, 2021_Aoy, 2022_Aoy}, spin excitations \cite{2021_Kat, 2024_Eto}, and dynamical properties \cite{2025_Shi}, thereby establishing it as a key topic in the broader field of topological magnetism.

\begin{figure*}[t]
\centering
\includegraphics[width=\linewidth]{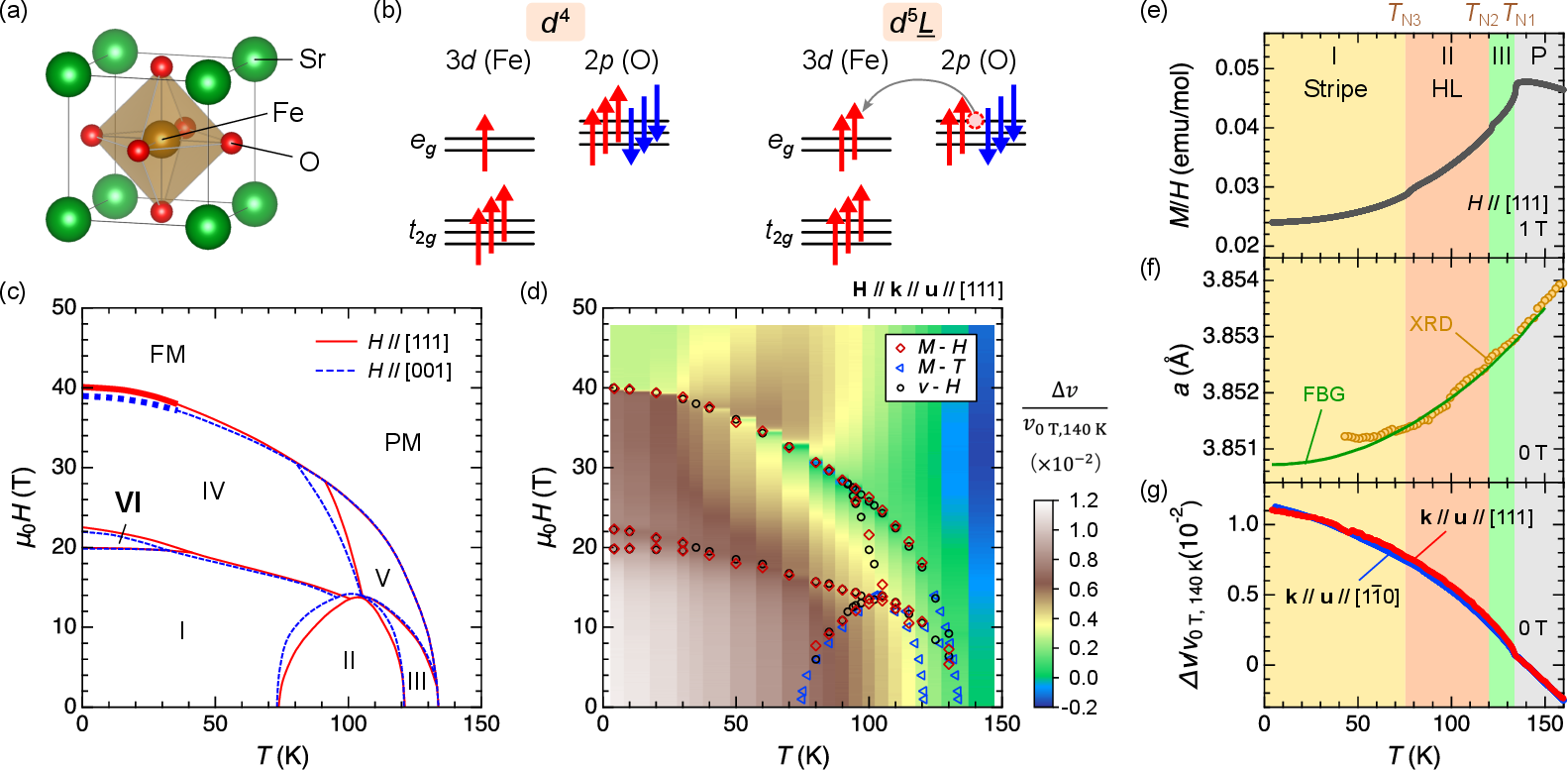}
\caption{(a) Crystal structure of SrFeO$_{3}$, illustrated using the VESTA software \cite{2011_Mom}. (b) Schematic electronic configurations of the $3d^{4}$ state (left) and the ligand-hole $3d^{5}{\underline L}$ state (right). (c) $H$--$T$ phase diagram for $H \parallel [111]$ (red solid line) and $H \parallel [001]$ (blue dashed line). The definitions of phases I--V follow Ref.~\cite{2011_Ish}, whereas phase VI is newly discovered in the present study. ``PM'' denotes the paramagnetic phase. The phase boundaries drawn with thick lines indicate a possible redistribution of the ligand-hole density from the $3d^{5}{\underline L}$ to the $3d^{4}$ state, as suggested by the anomalous magnetoelastic responses observed in this study. (d) Color map of the longitudinal sound velocity for ${\mathbf k} \parallel {\mathbf u} \parallel [111]$, normalized to its value at 0~T and 140~K, overlaid on the $H$--$T$ phase diagram for $H \parallel [111]$. (e) Temperature dependence of magnetic susceptibility $M/H$ at 1~T for $H \parallel [111]$. Three-step magnetic transitions are observed at $T_{\rm N1} = 134$~K, $T_{\rm N2} = 120$~K, and $T_{\rm N3} = 75$~K. (f) Temperature dependence of the cubic lattice constant $a$ at zero field, obtained from single-crystal synchrotron x-ray diffraction and thermal expansion measurements using the fiber Bragg grating (FBG) method. (g) Temperature dependence of the longitudinal sound velocity for ${\mathbf k} \parallel {\mathbf u} \parallel [111]$ (red) and ${\mathbf k} \parallel {\mathbf u} \parallel [1{\overline 1}0]$ (blue) in zero field.}
\label{Fig1}
\end{figure*}

The itinerant magnet SrFeO$_{3}$ has recently attracted considerable attention as a rare example of a centrosymmetric system hosting a quadruple-${\mathbf Q}$ HL \cite{1965_Mac, 1972_Tak, 1992_Boc, 2001_Hay, 2002_Abb,  2002_Nas, 2004_Leb, 2011_Ish, 2012_Lon, 2013_Cha, 2019_Rog, 2020_Ish, 2023_Kit, 2024_Tak, 2025_And}.
Despite its simple cubic perovskite structure [Fig.~\ref{Fig1}(a)], SrFeO$_{3}$ exhibits a rich variety of helimagnetic phases under high magnetic fields up to 40~T \cite{2011_Ish, 2012_Lon, 2013_Cha, 2020_Ish, 2025_And}.
Understanding the microscopic origin of the helimagnetic order in SrFeO$_{3}$ requires close attention to its unusual electronic properties.
The electronic configuration is characterized by the anomalously high Fe$^{4+}$ valence state, where the high-spin $3d^{4}$ configuration leaves the $e_{g}$ orbitals degenerate.
Because of the large negative charge-transfer energy \cite{1992_Boc, 2024_Tak}, the valence state is more appropriately described as a ligand-hole $3d^{5}{\underline L}$ configuration, where ${\underline L}$ denotes a hole in the oxygen $2p$ orbitals [Fig.~\ref{Fig1}(b)] \cite{2023_Kit}.
The formation of an oxygen-hole band preserves metallic conductivity down to low temperatures \cite{1965_Mac, 2011_Ish} and suppresses cooperative Jahn--Teller distortions \cite{2002_Abb, 2023_Kit}.
Magnetism is governed predominantly by the double exchange mechanism between Fe ions.
While the double exchange interaction generally favors a ferromagnetic (FM) state, Mostovoy {\it et al}. \cite{2005_Mos, 2017_Azh} theoretically proposed that the combination of negative charge-transfer energy and oxygen--oxygen hopping can stabilize incommensurate helical order, a scenario further supported by first-principles calculations \cite{2012_Li}.
Experimentally, the emergence of a FM state in SrFeO$_{3}$ was observed by high-pressure M\"{o}ssbauer spectroscopy above 13~GPa \cite{2002_Nas}.
The isovalent analogue CaFeO$_{3}$ undergoes a metal-insulator transition at 290 K, accompanied by charge disproportionation and breathing lattice distortion \cite{2000_Woo, 2018_Rog}.
Electron doping via La substitution in SrFeO$_{3}$ suppresses the multiple-${\mathbf Q}$ phases and induces coupled spin/charge order \cite{1997_Li, 2020_Ono}.
These findings demonstrate the rich interplay among spin, lattice, and charge degrees of freedom in iron oxides, highlighting the crucial role of Fe--O covalency and charge instability in determining both transport and magnetic properties in SrFeO$_{3}$. 

It has been well established that effective spin Hamiltonians incorporating momentum-resolved exchange interactions derived from the Kondo lattice model provide a powerful theoretical framework for describing multiple-${\mathbf Q}$ states in itinerant magnets \cite{2024_Eto, 2008_Mar, 2012_Aka, 2017_Oza, 2017_Hay, 2024_Hay}.
In centrosymmetric cubic lattices, not only biquadratic interactions but also anisotropic exchange interactions have been proposed as possible mechanisms stabilizing the HL phase \cite{2022_Kat}.
However, experimental knowledge of magnetic anisotropy in SrFeO$_{3}$ remains limited, as most previous studies have focused on magnetic fields applied along the [111] axis \cite{2011_Ish, 2012_Lon, 2020_Ish, 2025_And}.
Furthermore, the valence instabilities under applied magnetic fields remain unexplored, even though they may provide important clues to the stability of helimagnetic phases.
Magnetostriction and ultrasound measurements serve as powerful probes of field-induced valence fluctuations or valence transitions \cite{1999_Zhe, 2004_Mus, 2006_Mat, 2020_Kur, 2022_Miy, 2023_Nak}, yet systematic investigations of the magnetoelastic properties of SrFeO$_{3}$ are lacking.

Here, we revisit the magnetic phase diagram of SrFeO$_{3}$ by means of magnetization, magnetostriction, and ultrasound measurements in pulsed high-magnetic fields up to $\sim$50~T.
Our experimental results provide several novel insights: (i) The presence of an intermediate-field phase (phase~VI), overlooked in previous studies \cite{2011_Ish, 2025_And}, is revealed below 30~K. (ii) The saturation field is slightly lower for $H \parallel [001]$ than for $H \parallel [111]$, while the HL phase (phase~II) exhibits enhanced relative stability for $H \parallel [001]$ [Fig.~\ref{Fig1}(c)]. (iii) The magnetic transition to the forced FM phase is accompanied by a sharp metamagnetic jump, negative volume magnetostriction, and pronounced elastic softening below 35~K [Fig.~\ref{Fig1}(d)].
The latter observations would point to metal-to-ligand charge transfer accompanied by a reduction in the ligand-hole density, corresponding to a redistribution of electronic weight from the $3d^{5}\underline{L}$ to the $3d^{4}$ state.
In addition, we performed neutron scattering experiments in static magnetic fields tilted from [111] up to 17~T, which reveal domain selection rules in phase I (double-${\mathbf Q}$), single-${\mathbf Q}$ nature of phase~IV, and multiple-${\mathbf Q}$ nature of phase~V.
We theoretically show that the anisotropic phase diagram of SrFeO$_{3}$ can be qualitatively accounted for by an effective spin Hamiltonian incorporating a cubic single-ion anisotropy as well as bilinear and biquadratic interactions in momentum space.

\begin{figure*}[t]
\centering
\includegraphics[width=\linewidth]{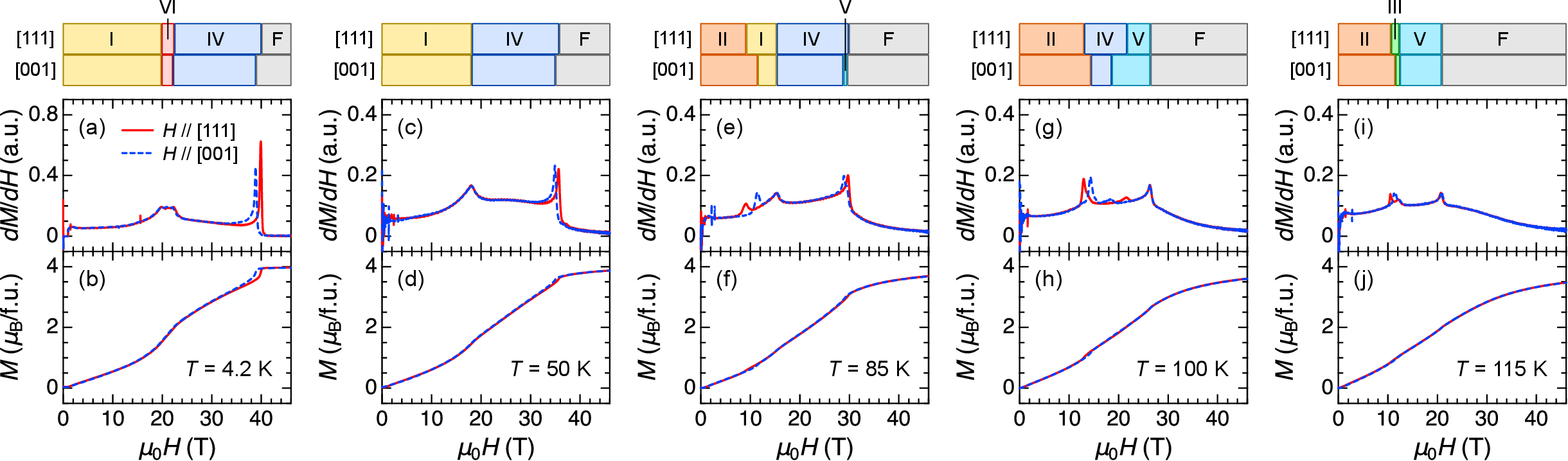}
\caption{Magnetic-field dependence of $M$ (bottom) and $dM/dH$ (top) at 4.2~K (a, b), 50~K (c, d), 85~K (e, f), 100~K (g, h), and 115~K (i, j) for $H \parallel [111]$ (red solid line) and $H \parallel [001]$ (blue dashed line). Only data for the field-increasing process are shown. The top bars indicate the corresponding magnetic phase diagrams, where ``F'' denotes the forced FM (or paramagnetic) phase.}
\label{Fig2}
\end{figure*}

\vspace{-0.3cm}
\section{\label{Sec2} Experimental results}

\vspace{-0.2cm}
\subsection{\label{Sec2-1} Magnetic and structural properties in zero field}
\vspace{-0.3cm}

SrFeO$_{3}$ crystallizes in a cubic perovskite structure with the space group $Pm{\overline 3}m$ (No.~225) at room temperature.
In zero magnetic field, SrFeO$_{3}$ undergoes successive magnetic transitions at $T_{\rm N1} = 134$~K, $T_{\rm N2} = 120$~K, and $T_{\rm N3} = 75$~K [Fig.~\ref{Fig1}(e)].
A previous neutron scattering study \cite{2020_Ish} proposed that phase I (below $T_{\rm N3}$) corresponds to an anisotropic double-${\mathbf Q}$ stripe phase, while phase II (between $T_{\rm N3}$ and $T_{\rm N2}$) corresponds to the quadruple-${\mathbf Q}$ HL phase.
These magnetic structures are characterized by superpositions of the modulation vector ${\mathbf Q}_{1} = (q, q, q)$ with $q \approx 0.13$ in reciprocal lattice units and its symmetry-equivalent counterparts \cite{1972_Tak, 2020_Ish}.

Although the magnetic structure in phase I breaks the cubic symmetry, our synchrotron x-ray diffraction and thermal expansion measurements reveal no detectable lattice-symmetry lowering, but instead only smooth changes in the lattice constants down to the lowest temperature [Fig.~\ref{Fig1}(f)].
These observations suggest that the lattice distortion associated with the magnetic long-range ordering is extremely small ($\Delta L/L < 10^{-4}$).
Furthermore, anomalies in the longitudinal sound velocity at zero field are observed only at $T_{\rm N1}$, followed by a monotonic hardening down to the lowest temperatures without any pronounced anisotropy in the propagation direction [Fig.~\ref{Fig1}(g)]. 
We consider that the three-dimensional character of the magnetic structures, together with the formation of the ligand-hole state, is essential for maintaining the structural stability without the cooperative Jahn--Teller distortion and giving rise to the relatively isotropic magnetoelastic coupling.

\vspace{-0.3cm}
\subsection{Field orientation dependence of magnetization curves}
\vspace{-0.3cm}

To extract information on the magnetic anisotropy, we measured the magnetization in pulsed high magnetic fields for two different field orientations, $H \parallel [111]$ and $H \parallel [001]$.
Figure~\ref{Fig2} summarizes the magnetization curves (bottom) and their field derivatives (top) during the field-increasing process for $H \parallel [111]$ (red solid lines) and $H \parallel [001]$ (blue dashed lines) at selected temperatures.
All magnetization data for $H \parallel [111]$ are presented in Fig.~\ref{Fig3}(a), and those for $H \parallel [001]$ are shown in Fig.~S1 in the Supplemental Material \cite{SM}.
Although the $H$--$T$ phase diagrams for $H \parallel [111]$ and $H \parallel [001]$ are qualitatively similar, several differences in the locations of the phase boundaries are resolved, as shown in Fig.~\ref{Fig1}(c).

At 4.2~K, the magnetization curves exhibit saturation behavior in the high-field region, where the magnetization $M$ reaches approximately 4~$\mu_{\rm B}$/f.u., consistent with the expected full moment for the Fe$^{4+}$ ion [Figs.~\ref{Fig2}(a) and \ref{Fig2}(b)].
The saturation field for $H \parallel [001]$, $\mu_{\rm 0}H_{\rm sat}^{[001]} = 39$~T, is slightly lower than that for $H \parallel [111]$, $\mu_{\rm 0}H_{\rm sat}^{[111]} = 40$~T.
This finding allows us to determine the sign of the coefficient in the cubic single-ion anisotropy term in the effective spin Hamiltonian, as discussed below.
Notably, the metamagnetic transition from phase IV to the forced FM phase is sharp and accompanied by hysteresis [see the inset in Fig.~\ref{Fig3}(a)], indicating its first-order nature.
Furthermore, we observe a double-hump anomaly in $dM/dH$ between 20 and 22~T [Fig.~\ref{Fig2}(a)], indicating the presence of an intermediate-field phase (phase VI).
These features were overlooked or not clearly observed in previous studies \cite{2011_Ish, 2025_And}.

As the temperature increases, phase VI disappears above 40~K for both field orientations [Figs.~\ref{Fig3}(a) and S1 \cite{SM}].
The difference in the phase diagram between $H \parallel [111]$ and $H \parallel [001]$ becomes more pronounced above 70~K [Figs.~\ref{Fig2}(e)--\ref{Fig2}(j)]: for $H \parallel [001]$, the boundaries between phases I and II, as well as between phases II and III, extend to higher fields, indicating enhanced stability of the quadruple-${\mathbf Q}$ HL phase.
In addition, the boundary between phases IV and V shifts to lower temperatures for $H \parallel [001]$.
As discussed below, the magnetic structures in phases~IV and V are single-${\mathbf Q}$ and multiple-${\mathbf Q}$, respectively, as revealed by our neutron scattering experiments.
These trends suggest that, under magnetic fields, the multiple-${\mathbf Q}$ order is more readily stabilized for $H \parallel [001]$ than for $H \parallel [111]$.

\begin{figure*}[t]
\centering
\includegraphics[width=\linewidth]{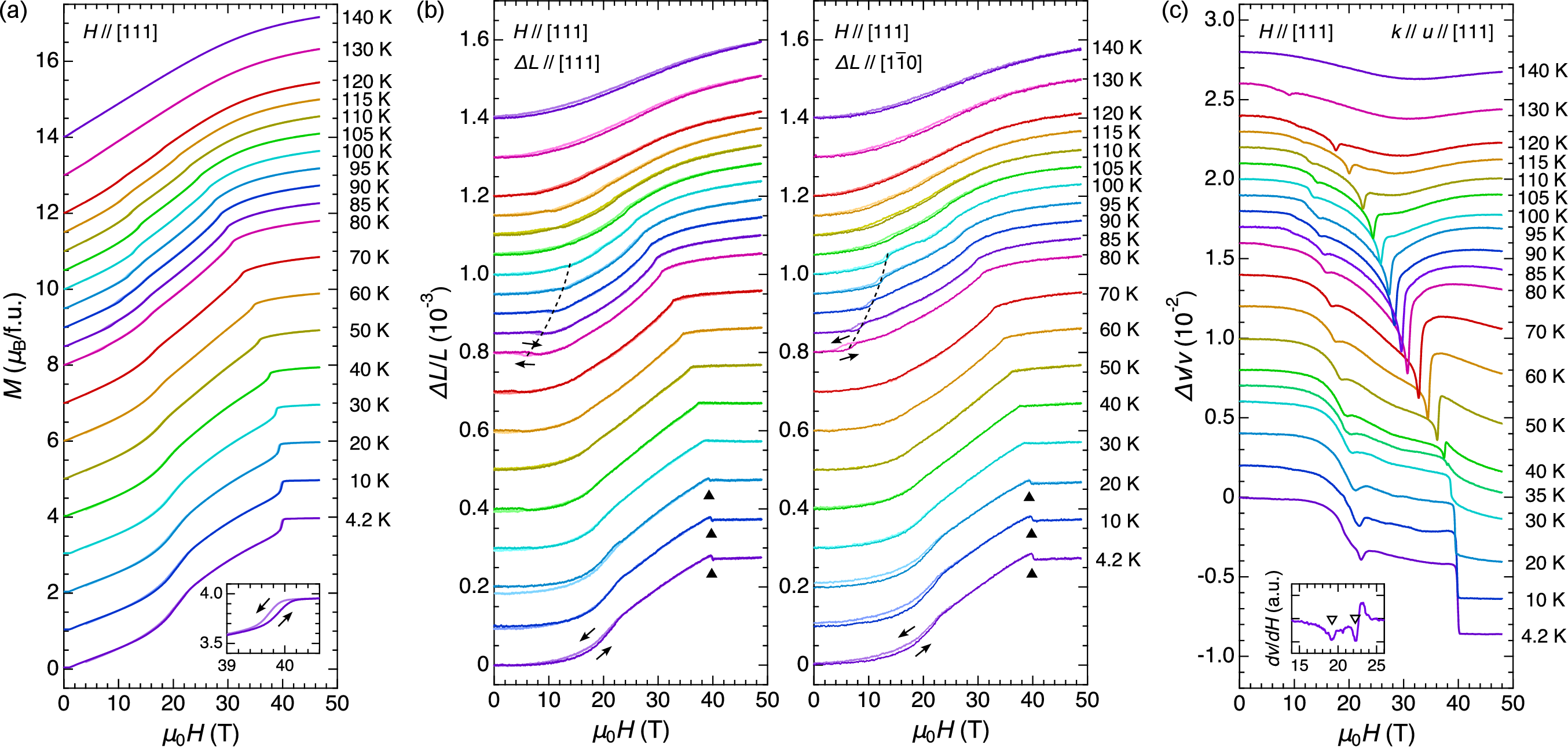}
\caption{Magnetic-field dependence of magnetization (a), longitudinal (left) and transverse (right) magnetostriction (b), and the longitudinal sound velocity for ${\mathbf k} \parallel {\mathbf u} \parallel [111]$ (c) measured at various temperatures for $H \parallel [111]$. For clarity, the curves are vertically shifted except for $T = 4.2$~K. In panels (a) and (b), data for the field-increasing (decreasing) process are shown in dark (light) colors. Only data for the field-decreasing process are shown in panel (c). The inset of panel (a) shows an enlarged view of the $M$--$H$ curve at 4.2~K near the saturation field. The inset of panel (c) shows the field-derivative of the sound velocity at 4.2~K, exhibiting a double-dip structure (denoted by triangles) around 20--22~T, which indicates the existence of phase VI.}
\label{Fig3}
\end{figure*}

\vspace{-0.3cm}
\subsection{Magnetoelastic properties and possible valence transition}
\vspace{-0.3cm}

We next move on to the magnetoelastic properties of SrFeO$_{3}$ at high magnetic fields.
Figure~\ref{Fig3}(b) shows the field dependence of the longitudinal (left) and transverse (right) magnetostriction for $H \parallel [111]$ at various temperatures, measured simultaneously on the same single crystal.
Both the longitudinal and transverse magnetostriction exhibit an overall isotropic expansion, with clear anomalies associated with the magnetic transitions.
Similar behaviors are also observed for $H \parallel [001]$ (Fig.~S2 \cite{SM}).
The relative change in length prior to saturation reaches $\Delta L/L \approx 2.8 \times 10^{-4}$ at 4.2~K, which is approximately three times larger than that reported in Ref.~\cite{2025_And}.
As indicated by the black dashed lines in Fig.~\ref{Fig3}(b), anisotropic magnetostriction is observed at the transition from phase II to phase I in the temperature range of 80--100~K.
Given the high symmetric quadruple-${\mathbf Q}$ HL state in phase II, this behavior suggests a slight lowering of lattice symmetry in phase I, presumably induced by the anisotropic double-${\mathbf Q}$ state.

\begin{figure*}[t]
\centering
\includegraphics[width=\linewidth]{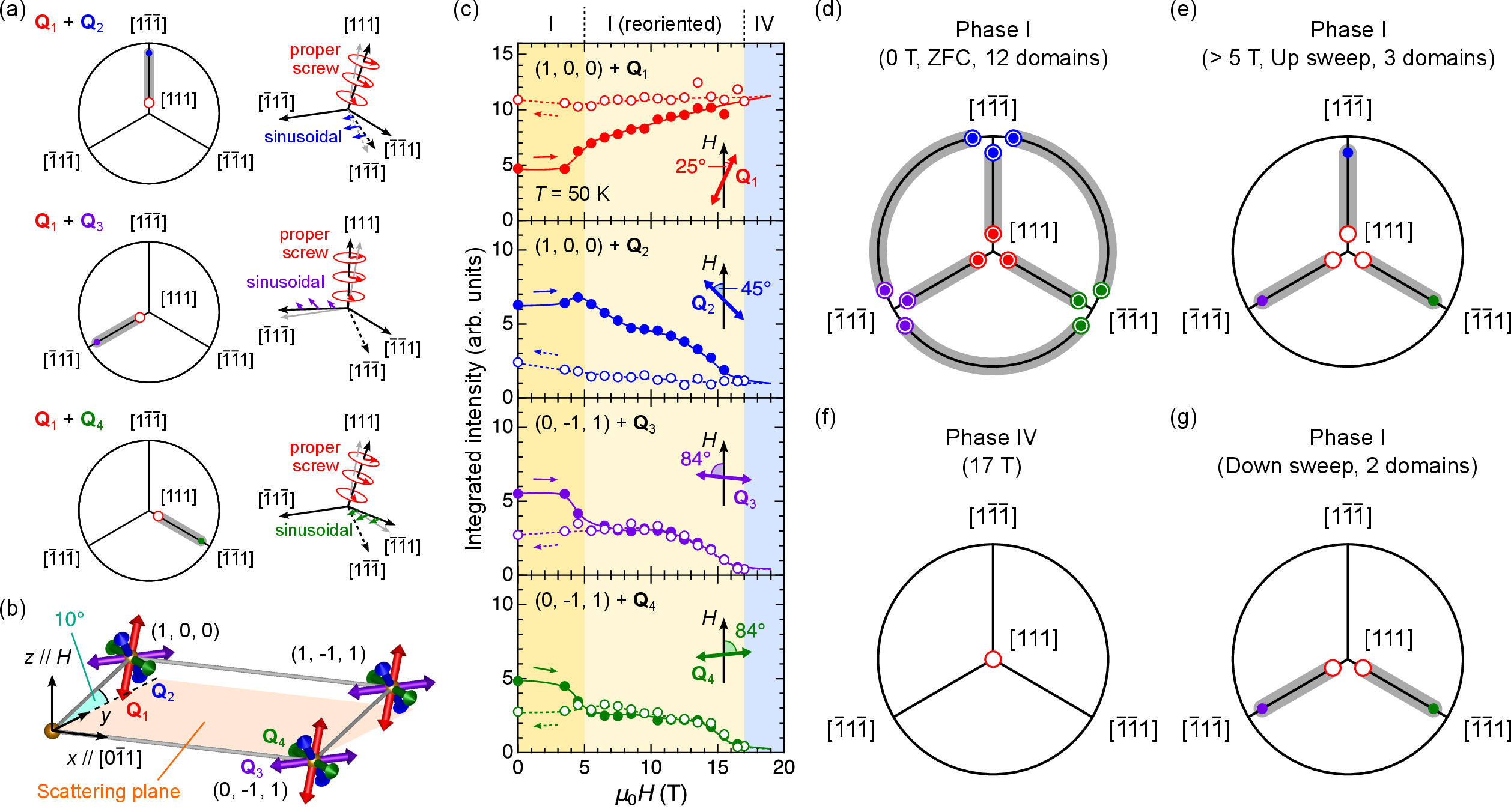}
\caption{(a) Three types of ${\mathbf Q}$-dependent domains in phase I that can be stabilized by applying a magnetic field along (near) the [111] axis. Although the neutron data alone do not uniquely distinguish between the sinusoidal and cycloidal modulations for the ${\mathbf Q}_{2}$--${\mathbf Q}_{4}$ components, theoretical calculations predicted the emergence of a double-${\mathbf Q}$ stripe phase composed of proper-screw and sinusoidal modulations. Each left panel shows stereographic projections of magnetic scattering peaks, where the filled and open circles represent magnetic reflections of the proper-screw and sinusoidal modulation, respectively. Each ${\mathbf Q}$ vector is slightly deviated from the $\langle 111 \rangle$ axis. (b) Schematic illustration of the geometric configuration in the neutron scattering experiment. A vertical magnetic field of up to 17~T was applied with a tilt of 25$^{\circ}$ from the [111] axis toward the $[100]$ axis. (c) Magnetic-field dependence of the integrated intensities of the ${\mathbf Q}_{1}$--${\mathbf Q}_{4}$ peaks measured at 50~K after zero-field cooling (ZFC). Data for the field-increasing (decreasing) process are plotted as filled (open) circles. The angle between each magnetic reflection vector and the applied field is shown in the inset of each panel. (d)--(g) Stereographic projections of magnetic scattering peaks at 0~T after ZFC (d), during the field-increasing process between 5 and 17~T (e), at 17~T in phase IV (f), and during the field-decreasing process (g). In panel (d), open circles with cores represent overlapping magnetic reflections of the proper-screw and sinusoidal types.}
\label{Fig4}
\end{figure*}

The most remarkable observation is the negative magnetostriction jump upon entering the forced FM phase below 30~K, detected in both longitudinal and transverse components, as indicated by the black triangles in Fig.~\ref{Fig3}(b).
This provides direct evidence for a negative volume effect, which cannot be explained within the framework of the conventional exchange striction mechanism or the band Jahn--Teller effect, and therefore points to the change in the Fe valence state.
Previous X-ray absorption spectroscopy \cite{2002_Abb} and valence electron density analysis \cite{2023_Kit} estimated the number of valence electrons at the Fe site to be 4.6--4.7.
The observed negative volume effect suggests an increased population of the Fe $3d^{4}$ state relative to the $3d^{5}{\underline L}$ state.
This is likely driven by charge transfer from the Fe~$3d$ to the O~$2p$ orbitals, whereby the system gains Madelung energy through a shortening of the Fe--O bonds.
This scenario can be understood within the theoretical framework of the Kondo-lattice model with double-exchange interactions proposed by Mostovoy \cite{2005_Mos}.
In this model, the ground state is predicted to be FM when the charge-transfer energy is positive, whereas a helimagnetic state is stabilized when it is negative.
Therefore, the present first-order transition into the forced FM state can be attributed to the suppression of the negative charge-transfer energy, which favors a FM state and allows the system to gain exchange energy.

Figure~\ref{Fig3}(c) shows the field dependence of the relative change in the longitudinal sound velocity $\Delta v/v$ for propagation along ${\mathbf k} \parallel {\mathbf u} \parallel [111]$ under $H \parallel [111]$.
The sound velocity normalized to its value at 0~T and 140~K is displayed as a color plot in the $H$--$T$ phase diagram in Fig.~\ref{Fig1}(d).
All magnetic transitions are clearly detected as distinct anomalies in both the sound velocity and attenuation (Fig.~S3 \cite{SM}).
The presence of phase VI is confirmed by the double-dip anomaly in $dv/dH$ around 20~T, as indicated by the open triangles in the inset of Fig.~\ref{Fig3}(c).
The possible change in the Fe valence state is further indicated by the characteristic $\Delta v/v$ behavior at high fields: a peak-shaped anomaly is observed across the transition into the forced FM phase above 35~K, whereas a sharp step-like anomaly appears below 35~K.
The dramatic softening of the sound velocity, despite the negative volume expansion, points to a structural instability in the forced FM state below 35~K, presumably induced by the increased population of the $3d^{4}$ state.
Considering that similar ultrasonic responses are observed for the ${\mathbf k} \parallel {\mathbf u} \parallel [111]$ and ${\mathbf k} \parallel {\mathbf u} \parallel [1\overline{1}0]$ geometries (Fig.~S3 \cite{SM}), this structural instability is unlikely to be related to a Jahn--Teller instability toward tetrahedral distortion arising from the lifting of the $e_g$ orbital degeneracy.
Rather, we infer that the structural instability is associated with an electronic instability toward charge disproportionation, which is typically accompanied by breathing lattice distortion, as reported in CaFeO$_{3}$ \cite{2000_Woo, 2018_Rog} and Sr$_{3}$Fe$_{2}$O$_{7}$ \cite{2000_Kuz, 2021_Kim}.

\vspace{-0.3cm}
\subsection{Neutron scattering experiments under tilted magnetic fields}
\vspace{-0.3cm}

The multiple-${\mathbf Q}$ nature in phases I and II was confirmed by a previous single-crystal neutron scattering experiment under $H \parallel [111]$ \cite{2020_Ish}.
In both phases, the authors observed four magnetic Bragg reflections corresponding to ${\mathbf Q}_{1} = (q, q, q)$, ${\mathbf Q}_{2} = (q, -q, -q)$, ${\mathbf Q}_{3} = (-q, q, -q)$, and ${\mathbf Q}_{4} = (-q, -q, q)$, where $q \approx 0.13$~r.l.u.
Based on the field dependence of the intensity distribution after zero-field cooling (ZFC), phase I was proposed to be an anisotropic double-${\mathbf Q}$ state, whereas phase II was identified as an isotropic quadruple-${\mathbf Q}$ state.
Polarized neutron scattering further suggested that phase I consists of a superposition of a proper-screw and a sinusoidal (or cycloidal) modulation \cite{2020_Ish}, allowing for twelve ${\mathbf Q}$-dependent domains.
The application of a magnetic field along [111] would stabilize three of them, consisting of the proper-screw component of ${\mathbf Q}_{1}$ and the sinusoidal (or cycloidal) component of ${\mathbf Q}_{i}$ ($i = 2, 3, 4$), as shown in Fig.~\ref{Fig4}(a).
In addition, small-angle neutron scattering revealed triple-peak splitting of the Bragg spot in phase I, indicating slight deviations of the modulation vectors from the high-symmetric $\langle 111 \rangle$ axes, e.g., ${\mathbf Q}_{1} = (q', q, q)$ and ${\mathbf Q}_{2} = (q', -q, -q)$ for the ${\mathbf Q}_{1}+{\mathbf Q}_{2}$ domain, where $q' > q$ \cite{2020_Ish}.

\begin{figure*}[t]
\centering
\includegraphics[width=\linewidth]{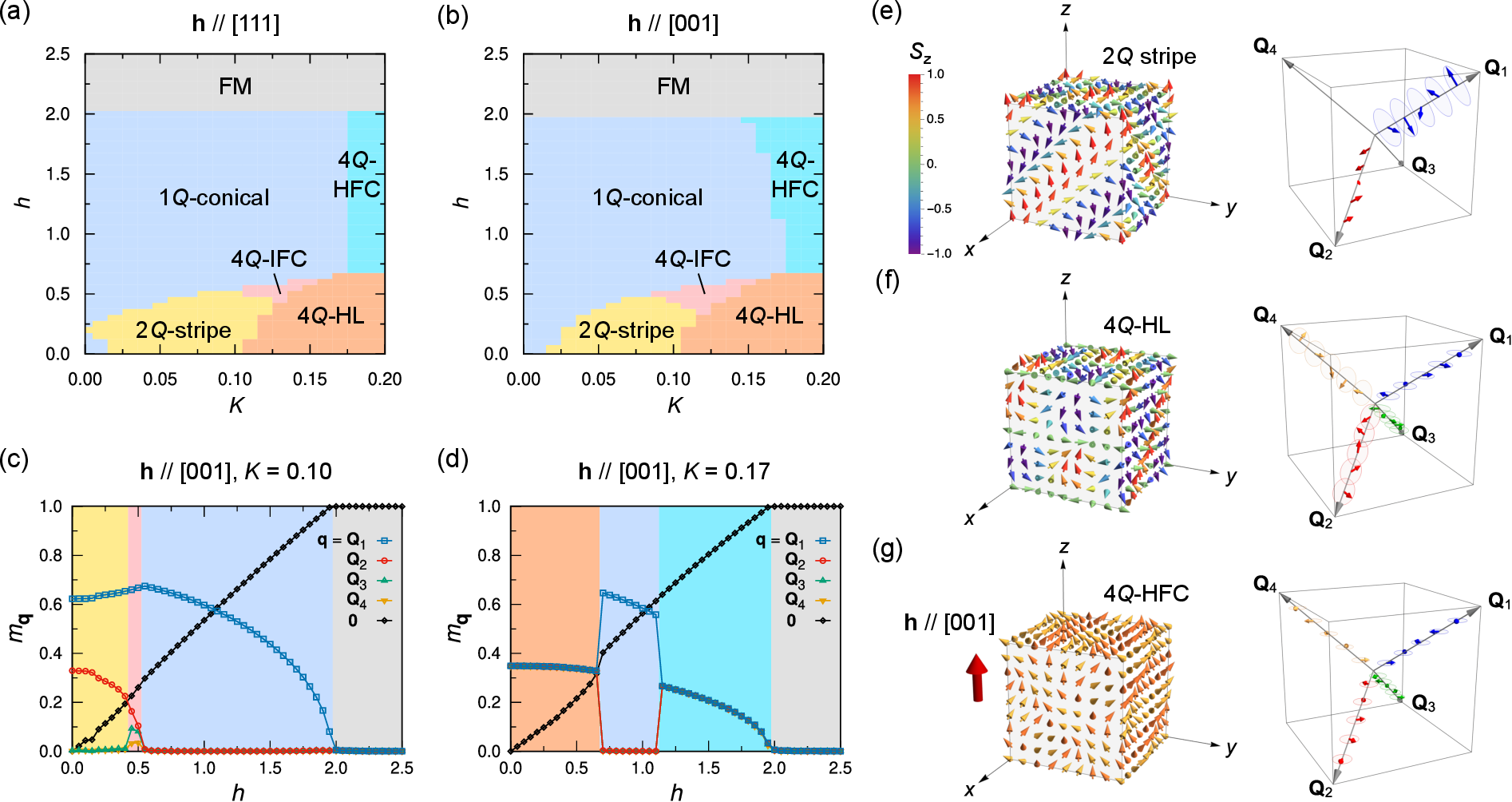}
\caption{(a, b) Ground-state phase diagrams of the effective bilinear--biquadratic model Eq.~\eqref{Eq1} with $A = -0.01$ as functions of $K$ and $h$ for ${\mathbf h} \parallel [111]$ (a) and ${\mathbf h} \parallel [001]$ (b). (c, d) Magnetic-field dependence of the magnetic moments with wave vectors, $m_{\mathbf q}$, for $K=0.10$ (c) and $K=0.17$ (d) for ${\mathbf h} \parallel [001]$ (d). $m_{{\mathbf q} = 0}$ represents the magnetization. The background colors indicate different magnetic phases, following the color scheme used in the phase diagram in panel (b). (e)--(g) Spin textures (left) and constituent waves (right) for 2{\it Q}-stripe (e), 4{\it Q}-HL (f), and 4{\it Q}-HFC phases (g). The color bar represents the spin $z$ component, which is parallel to the field direction.}
\label{Fig5}
\end{figure*}

However, the previous neutron scattering did not observe an explicit breaking of the threefold rotational symmetry in phase~I, and therefore the double-${\mathbf Q}$ nature could not be established unambiguously.
Although a subsequent theoretical study \cite{2016_Oza, 2022_Oku} proposed the emergence of an anisotropic double-${\mathbf Q}$ order that is consistent with the experimental proposal \cite{2020_Ish}, theory also allows for alternative scenarios, such as anisotropic quadruple-${\mathbf Q}$ states that preserve threefold symmetry.
Therefore, it remained necessary to experimentally discriminate between these competing possibilities and to establish the magnetic structure of phase~I.
To this end, we performed single-crystal neutron scattering experiments under high magnetic fields up to 17~T, applied with a tilt of $25^{\circ}$ from the [111] axis toward the [100] axis.
In this configuration, magnetic reflections corresponding to ${\mathbf Q}_{1}$ and ${\mathbf Q}_{2}$ around the $(1, 0, 0)$ reciprocal lattice point, as well as ${\mathbf Q}_{3}$ and ${\mathbf Q}_{4}$ around $(0, -1, 1)$, were observed within a deviation of less than $2^{\circ}$ from the horizontal scattering plane [Fig.~\ref{Fig4}(b)].
In contrast to the previous study \cite{2020_Ish}, the magnetic field breaks the threefold symmetry with respect to the [111] axis, rendering the ${\mathbf Q}_{2}$ vector inequivalent to the ${\mathbf Q}_{3}$ and ${\mathbf Q}_{4}$ vectors.

Figure~\ref{Fig4}(c) shows the field dependence of the integrated intensities of the ${\mathbf Q}_{1}$--${\mathbf Q}_{4}$ peaks, measured at 50~K after ZFC.
We note that the expected triple-peak splitting \cite{2020_Ish}, as mentioned above, was observed as a single broadened peak due to the limited experimental resolution (Fig.~S4 \cite{SM}).
The initial state at 0~T is expected to contain twelve magnetic domains of the double-${\mathbf Q}$ state [Fig~\ref{Fig4}(d)], as suggested by the nearly equal intensities across the four peaks.
Upon increasing the magnetic field beyond 5~T, the intensity of the ${\mathbf Q}_{1}$ peak increases, whereas those of the other peaks decrease but remain finite, consistent with the previous study \cite{2020_Ish}.
This behavior suggests a domain reorientation into three domains with the proper-screw ${\mathbf Q}_{1}$ component [Figs.~\ref{Fig4}(a) and \ref{Fig4}(e)].
At 17~T, the ${\mathbf Q}_{2}$--${\mathbf Q}_{4}$ peaks nearly vanish, indicating the emergence of a single-${\mathbf Q}$ proper-screw state in phase~IV [Fig.~\ref{Fig4}(f)].
During the subsequent field-decreasing process, the ${\mathbf Q}_{3}$ and ${\mathbf Q}_{4}$ peaks rapidly recover their intensities, whereas the ${\mathbf Q}_{2}$ peak remains nearly absent, suggesting that the ${\mathbf Q}_{1}+{\mathbf Q}_{3}$ and ${\mathbf Q}_{1}+{\mathbf Q}_{4}$ domains are energetically favored over the ${\mathbf Q}_{1}+{\mathbf Q}_{2}$ domain.
The observed three-fold symmetry breaking offers direct evidence for the double-${\mathbf Q}$ nature in phase~I.

We also performed high-field neutron scattering at 110 K (Fig.~S5 \cite{SM}).
In contrast to the measurements at 50 K, the intensities of the ${\mathbf Q}_{2}$--${\mathbf Q}_{4}$ peaks do not completely vanish even after the field-induced phase transition, suggesting a multiple-${\mathbf Q}$ nature of phase~V.

\vspace{-0.3cm}
\section{\label{Sec3} Theoretical calculations}
\vspace{-0.3cm}

For a deeper understanding of the $H$--$T$ phase diagram of SrFeO$_{3}$ [Fig.~\ref{Fig1}(c)], we consider the effective spin model on a simple cubic lattice \cite{2022_Oku}, incorporating long-range exchange interactions and local terms, which is given by
\begin{equation}
\begin{split}
\label{Eq1}
{\mathcal{H}_{\rm eff}} =&~2\sum_{\eta}\left[-J{\mathbf S}_{{\mathbf Q}_{\eta}}\cdot{\mathbf S}_{-{\mathbf Q}_{\eta}}+\frac{K}{N}\left({\mathbf S}_{{\mathbf Q}_{\eta}}\cdot{\mathbf S}_{-{\mathbf Q}_{\eta}}\right)^{2}\right]\\&+A\sum_{l}\left[(S_{l}^{x})^4+(S_{l}^{y})^4+(S_{l}^{z})^4\right]-\sum_{l}{\mathbf h}\cdot{\mathbf S}_{l},
\end{split}
\end{equation}
where ${\mathbf S}_{\mathbf q}=\frac{1}{\sqrt{N}}\sum_{l}{\mathbf S}_{l}e^{i{\mathbf q}\cdot{\mathbf r}_{l}}$ is the Fourier transform of the localized moment ${\mathbf S}_{l}=(S_{l}^{x}, S_{l}^{y}, S_{l}^{z})$ at a position ${\mathbf r}_{l}$, and $N$ is the number of spins.
The first term represents the multispin exchange interactions in momentum space, comprising the bilinear and biquadratic interactions, which are featured by characteristic wavenumbers in the electronic structure \cite{2012_Aka, 2017_Hay}.
The bilinear interaction stabilizes a spin modulation with the wavenumber ${\mathbf Q}_{\eta}$ and the energy scale is set to $J = 1$.
The biquadratic interaction with a positive coupling constant $K > 0$, which is the most relevant term among the higher-order perturbations with respect to the spin--charge coupling, allows a superposition of several wavenumbers.
The second term represents a cubic single-ion anisotropy, where $A>0$ ($A<0$) corresponds to a magnetic easy axis along [111] ([100]) and symmetrically equivalent directions. 
The last term describes the Zeeman coupling to an external magnetic field ${\mathbf h}$.

We numerically investigate the ground state of Eq.~\eqref{Eq1} by means of simulated annealing for a system of $N=8^{3}$ spins with periodic boundary conditions (see Appendix~\ref{AppendixA} for details).
Following the experimentally observed magnetic modulation vector in SrFeO$_{3}$, ${\mathbf Q} = (q, q, q)$ ($q \approx 0.13$~r.l.u) \cite{1972_Tak, 2020_Ish}, the summation in Eq.~\eqref{Eq1} is taken for the set of tetrahedral wave vectors, ${\mathbf Q}_{1} = (Q, Q, Q)$, ${\mathbf Q}_{2} = (Q, -Q, -Q)$, ${\mathbf Q}_{3} = (-Q, Q, -Q)$, and ${\mathbf Q}_{4} = (-Q, -Q, Q)$, where we set $Q = \pi/4$, corresponding to a period of eight lattice sites.
The slight deviation of the ${\mathbf Q}$ vectors from the high-symmetric $\langle 111 \rangle$ axis in phase I is ignored.
Figures~\ref{Fig5}(a) and \ref{Fig5}(b) show the theoretical ground-state phase diagrams with $A = -0.01$ as functions of $K$ and $h$ for magnetic fields applied along the [111] and [001] axes, i.e., ${\mathbf h} = \frac{1}{\sqrt{3}}(h, h, h) \parallel \mathbf{Q}_{1}$ and ${\mathbf h} = (0, 0, h)$, respectively.
By setting $A < 0$, the saturation field becomes slightly larger for ${\mathbf h} \parallel [111]$ than for ${\mathbf h} \parallel [001]$, in agreement with the experimental observations [Fig.~\ref{Fig1}(c)].

We first note that the anisotropic double-${\mathbf Q}$ state, which is the zero-field ground state of SrFeO$_{3}$, is well reproduced in the small $K$ regime, $0.02 \lesssim K \lesssim 0.10$.
Although the polarized neutron data alone do not uniquely distinguish between the cycloidal and sinusoidal components in phase~I \cite{2020_Ish}, our theoretical calculations propose the emergence of a double-${\mathbf Q}$ stripe state composed of proper-screw and sinusoidal modulations \cite{2016_Oza, 2022_Oku} [Fig.~\ref{Fig5}(e)].
Notably, upon applying a magnetic field to this double-${\mathbf Q}$ phase, successive phase transitions occur via an intermediate-field phase with a pronounced linear slope in the magnetization process, eventually leading to a single-${\mathbf Q}$ conical phase at high fields [Fig.~\ref{Fig5}(c)].
This high-field single-${\mathbf Q}$ phase can be naturally associated with phase~IV in SrFeO$_{3}$.
The intermediate-field phase is theoretically identified as a topologically trivial quadruple-${\mathbf Q}$ conical state (4{\it Q}-IFC), consisting of the mixture of helix and sinusoidal components (see Fig.~S6 for details \cite{SM}).
This may correspond to the newly discovered phase~VI in SrFeO$_{3}$, based on the similarity in the magnetization process [Fig.~\ref{Fig2}(b)].

We also propose that the ground-state phase diagrams in the larger $K$ regime can be regarded as effectively incorporating finite-temperature effects \cite{1991_Rei, 2011_Oku}, since the introduction of biquadratic interactions tends to stabilize multiple-${\mathbf Q}$ states and thereby increase the magnetic entropy.
Indeed, for $K \gtrsim 0.11$, a topologically nontrivial quadruple-${\mathbf Q}$ HL phase (4{\it Q}-HL) emerges as the zero-field state [Fig.~\ref{Fig5}(f)], hosting eight monopole-anti monopole pairs~\cite{2022_Oku}, while for even larger $K$, a topologically trivial quadruple-${\mathbf Q}$ conical state (4{\it Q}-HFC) composed of four helix components appears in the high-field region [Fig.~\ref{Fig5}(g)].
In light of the neutron scattering observation that phase~V exhibits a multiple-${\mathbf Q}$ nature (Fig.~S5 \cite{SM}), phase~V is likely to correspond to the 4{\it Q}-HFC phase.
Interestingly, for moderate values of $K$, a magnetic field induces a reentrant sequence of phase transitions, 4{\it Q}-HL $\rightarrow$ 1{\it Q}-conical $\rightarrow$ 4{\it Q}-HFC, where the quadruple-${\mathbf Q}$ order is suppressed at intermediate fields and restored at higher fields.
Furthermore, for all quadruple-${\mathbf Q}$ phases, the parameter regions in which they are stabilized are broader for ${\mathbf h} \parallel [001]$ than for ${\mathbf h} \parallel [111]$.
These trends capture the essential features of the $H$--$T$ phase diagram of SrFeO$_{3}$ [Fig.~\ref{Fig1}(c)], demonstrating that the present bilinear-biquadratic model provides a unified description of the field-induced phase transitions and that even a weak cubic single-ion anisotropy has a pronounced impact on the stability of multiple-${\mathbf Q}$ phases.

\vspace{-0.2cm}
\section{Discussion and conclusions}
\vspace{-0.2cm}

We have investigated the high-field phase diagram of SrFeO$_{3}$ through a combination of magnetization, magnetostriction, and ultrasound measurements under pulsed high-magnetic fields.
All phase transitions were clearly detected as magnetostrictive and ultrasonic anomalies, highlighting the substantial magnetoelastic coupling in this compound.
Furthermore, single-crystal neutron scattering experiments under tilted magnetic fields enabled us to clarify the magnetic modulation vectors in each phase: phase~I is a double-${\mathbf Q}$ state, phase~II is a quadruple-${\mathbf Q}$ state, phase~IV is a single-${\mathbf Q}$ state, and phase~V is a multiple-${\mathbf Q}$ state.
Remarkably, these experimental observations are well captured by an effective bilinear-biquadratic spin model.
Our theoretical calculations based on the bilinear-biquadratic model suggest that the double-${\mathbf Q}$ state in phase~I is characterized by a superposition of helix and sinusoidal components, and phases~V and VI may correspond to topologically trivial quadruple-${\mathbf Q}$ states, in contrast to the topologically nontrivial quadruple-${\mathbf Q}$ HL state realized in phase~II.
We also revealed the field-orientation dependence of the magnetic phase diagram,  showing that the saturation field is slightly lower for $H \parallel [001]$ than for $H \parallel [111]$, whereas the HL phase appears over a relatively broader $H$--$T$ regime for $H \parallel [001]$. 
These features can be explained by introducing a cubic single-ion anisotropy term that favors the [001] axis as the magnetic easy axis, suggesting that the stability of multiple-${\mathbf Q}$ phases is highly sensitive to magnetic anisotropy.

Beyond deepening the understanding of multiple-${\mathbf Q}$ physics on a simple cubic lattice, another important implication of the present work is the renewed insight into the origin of helimagnetism in SrFeO$_{3}$, highlighting the essential role of Fe--O covalency \cite{2005_Mos}.
The experimentally observed sharp metamagnetic transition immediately below the saturation [Fig.~\ref{Fig3}(a)] cannot be reproduced within the theoretical framework of the effective spin model Eq.~\eqref{Eq1} [Figs.~\ref{Fig5}(c) and \ref{Fig5}(d)], suggesting that additional electronic degrees of freedom, such as field-dependent Fe $3d$--O $2p$ hybridization, play an essential role in the metamagnetic transition.
Notably, signatures of metal-to-ligand charge transfer are also evident from the associated negative volume jump and pronounced elastic anomalies [Figs.~\ref{Fig3}(b) and \ref{Fig3}(c)].
If the ligand-hole density is assumed to remain unchanged throughout the entire field range, the magnetization in phase~IV is expected to increase smoothly toward saturation, yielding a hypothetical saturation field of approximately 45~T estimated from an extrapolation of the magnetization curve.

It should be noted that the observed negative volume change at $H_{\rm sat}^{[111]}$ is relatively small, amounting to $\Delta V/V \approx -5 \times 10^{-5}$.
Even when taking into account the possibility that the positive volume expansion associated with the increase in magnetization partially compensates this contribution, the intrinsic volume change associated with the Fe valence state change is still estimated to be as small as $\Delta V/V \approx -1 \times 10^{-4}$.
By contrast, valence transitions in Eu- and Yb-based compounds typically produce magnetostrictive changes on the order of $10^{-3}$ \cite{2004_Mus, 2006_Mat, 2023_Nak}.
This marked difference may reflect a distinct microscopic mechanism of the valence state change: while Eu- and Yb-based systems involve a change in the ionic valence of localized $4f$ electrons, the present case in SrFeO$_{3}$ is likely governed by a more subtle redistribution of ligand-hole density associated with Fe $3d$--O $2p$ hybridization.
We therefore infer that the variation in the Fe valence in SrFeO$_{3}$ is relatively small, and that the system remains close to the ligand-hole $3d^{5}\underline{L}$ configuration even in the forced FM phase.
Furthermore, we observe pronounced elastic softening at the FM transition, suggesting enhanced structural instability  associated with charge disproportionation \cite{2000_Woo, 2018_Rog, 2000_Kuz, 2021_Kim}.
Future spectroscopic studies under high magnetic fields, such as X-ray absorption spectroscopy, are highly desirable to directly clarify the change in the ligand-hole density as well as the Fe valence state associated with the metamagnetic transition.

In summary, we have substantially updated the high-field phase diagram of SrFeO$_{3}$ using magnetoelastic probes and established an effective theoretical framework that accounts for the key features of its field-induced phases.
Our results indicate that a FM state is hidden as a metastable ground state in SrFeO$_{3}$, and that the electronic itinerancy arising from the formation of the ligand-hole band is essential for stabilizing its diverse helical and multiple-${\mathbf Q}$ magnetic phases.
The determination of the detailed magnetic structure in a newly-discovered phase~VI remains an important subject for future work, which would be enabled by pulsed-field neutron scattering experiments \cite{2024_Nak}.
High-precision electrical transport measurements under high magnetic fields, particularly Hall-resistivity measurements, will provide an important avenue for exploring the interplay between multiple-${\mathbf Q}$ magnetism and emergent transport phenomena, such as the topological Hall effect, and for detecting field-induced changes in the ligand-hole density.

\vspace{-0.2cm}
\section*{Acknowledgments}
\vspace{-0.2cm}
This work was financially supported by Japan Society for the Promotion of Science (JSPS) KAKENHI Grant-in-Aid for Scientific Research (No.~20J10988, No.~23K13068, No.~24H01633, No.~25H00420, and No.~26H00585).
The synchrotron radiation experiments were performed at SPring-8 with the approval of the Japan Synchrotron Radiation Research Institute (JASRI) (Proposal No.~2022A1751).
The neutron scattering experiment at HZB was carried out along the proposals (No.~14201044-ST).
The authors thank H. Takahashi and Y. Kobayashi for supporting the high-pressure synthesis of SrFeO$_{3}$ single crystals, and A. Matsuo for technical support of the magnetization measurmeent in pulsed high-magnetic fiedd.

\appendix

\vspace{-0.2cm}
\section{\label{AppendixA} Methods}

Single crystals of SrFeO$_{3}$ were obtained by high-pressure oxygen annealing of large single crystals of the oxygen-deficient perovskite SrFeO$_{2.5}$ with a brownmillerite-type structure, as described in Ref.~\cite{2011_Ish}.
A single crystal of SrFeO$_{2.5}$ was grown by a floating-zone method in an argon gas flow.
The resulting cylindrical crystal, with a diameter of approximately 4~mm, was cut to fit a gold capsule and then treated with the oxidizer NaClO$_{3}$ for 1h at 873~K and 8~GPa.
We confirmed the high crystallinity and the absence of detectable oxygen deficiency in the obtained SrFeO$_{3}$ single crystals using high-resolution x-ray diffraction and x-ray absorption spectroscopy \cite{2023_Kit, 2024_Tak}.
The temperature dependence of the lattice constant was measured using a single-crystal x-ray diffractometer at BL02B1 of the synchrotron radiation facility SPring-8, Japan, with an incident x-ray energy of $E = 40$~keV.
We prepared two crystals from the same batch for physical property measurements in pulsed high magnetic fields, one for $H \parallel [111]$ and the other for $H \parallel [001]$.

Magnetization up to 7~T was measured using a SQUID magnetometer (MPMS, Quantum Design), and up to 14~T using a commercial vibrating sample magnetometer (PPMS, Quantum Design).
Magnetization up to 47~T was measured by the induction method in a nondestructive pulsed magnet with a 4 ms pulse duration.
The magnetic field was applied along [111] and [001].
The absolute values of magnetization were calibrated using data obtained in MPMS or PPMS.
Thermal expansion was measured along the  [111] axis at zero field by the fiber-Bragg-grating (FBG) method using an optical sensing instrument (Hyperion si155, LUNA) in PPMS.
Magnetostriction measurements were performed by the FBG method up to 49~T in a nondestructive pulsed magnet with a 36 ms pulse duration.
The magnetic field was applied along [111] and [001].
Longitudinal and transverse magnetostriction were simultaneously measured by attaching two optical fibers orthogonally to the top and side surfaces of a parallelepiped-shaped single crystal using Stycast 1266 epoxy.
The optical filter method was employed to detect the relative change in sample length~\cite{2018_Ike}.
Ultrasound measurements were performed using the standard pulse-echo method up to 48~T in a nondestructive pulsed magnet with a 36 ms pulse duration.
The magnetic field was applied along [111].
Longitudinal ultrasound waves with a frequency of 50~MHz were generated and detected by LiNbO$_{3}$ piezoelectric transducers (60 $\mu$m thickness) attached to the side surfaces of the crystal.
Measurements were carried out with two different configurations: ${\mathbf k} \parallel {\mathbf u} \parallel [111]$ and ${\mathbf k} \parallel {\mathbf u} \parallel [1{\overline 1}0]$, where ${\mathbf k}$ and ${\mathbf u}$ are propagation and polarization vectors, respectively.
Temperature dependence of the sound velocity in zero field was measured using the same setup.
All the experiments in pulsed high magnetic fields were performed at the Institute for Solid State Physics, University of Tokyo, Japan.

Neutron diffraction experiment on the single crystal SrFeO$_{3}$ was performed at the two-axis neutron diffractometer E4 at Helmholtz-Zentrum Berlin (HZB).
The sample was attached to an Al sample holder and loaded into the vertical-field superconducting cryomagnet VM-1 with the maximum field of 14.5~T.
To extend the field range, we employed the Dy-booster insert \cite{2001_pro}, which has two pole tips of highly textured polycrystalline Dy.
The sample was mounted between the two Dy poles.
By applying an external magnetic field larger than approximately 2~T at low temperatures, the Dy poles were fully magnetized, providing an additional magnetic field of 2.5~T at the sample position.
Then, the field range at the sample position was extended to 17~T.  
An incident neutron beam with a wavelength of 2.44~$\AA$ was obtained by a pyrolytic graphite (PG) monochromator.
The horizontal scattering plane is shown in Fig.~\ref{Fig4}(b).
The $[0{\overline 1}1]$ direction was selected to be parallel to the horizontal plane, and the [100] direction was tilted by 10 degrees from the horizontal plane.
This tilted scattering geometry enabled us to measure satellite reflections belonging to all the four ${\mathbf Q}$ vectors, specifically $(1,0,0)+{\mathbf Q}_{1}$, $(1,0,0)+{\mathbf Q}_{2}$, $(0, {\overline 1}, 1)+{\mathbf Q}_{3}$, and $(0, {\overline 1}, 1)+{\mathbf Q}_{4}$.
Although these reflections were not exactly located on the horizontal plane, the deviations were small enough to quantitatively measure the intensities of the scattered neutrons.

Simulated annealing is performed with gradually reducing temperature from $T=1$ to $T=10^{-5}$ under a condition $T_n=10^{-0.1n}$, where $T_n$ is the temperature in the $n$th step.
We spend a total of $10^5-10^6$ Monte Carlo sweeps during the annealing by using the standard Metropolis algorithm.
After annealing at a set of $K$ and $h$, we increase or decrease $K$ and $h$ successively by $\Delta K=0.01$ and $\Delta h= 0.05$, respectively.
At every shift by $\Delta K$ or $\Delta h$, we heat the system to $T=10^{-3}$ and cool it down again to $T=10^{-5}$ by the same scheme of annealing. 
Carefully comparing the results by starting from several initial states for various $K$ and $h$, we obtain the lowest-energy state at each $K$ and $h$. 
To identify the magnetic phases, we calculate the magnetic moment with wave vector $\mathbf{q}$, $m_\mathbf{q}=\sqrt{S(\mathbf{q})/N}$, where $S(\mathbf{q})$ is the spin structure factor defined by $S(\mathbf{q})=\frac{1}{N}\sum_{l,l'}\mathbf{S}_{l}\cdot\mathbf{S}_{l'}e^{i\mathbf{q}\cdot(\mathbf{r}_{l}-\mathbf{r}_{l'})}$; $m_{\mathbf{q}=0}$ corresponds to the magnetization per spin.

\end{document}